\documentstyle[prl,aps,twocolumn,epsf]{revtex}
\newcommand{\r}{{\bf r}}
\begin{document}
\title{\large 
{\bf Magnetization of mesoscopic disordered networks}}
 \author{M. Pascaud and G.
Montambaux} \address{Laboratoire de Physique des Solides,  associ\'e au
CNRS \\ Universit\'{e} Paris--Sud \\ 91405 Orsay, France}
\twocolumn[
\date{\today}
\maketitle
\widetext
 \begin{center}
 \begin{abstract}
\parbox{14cm}{
We study the magnetic response of
mesoscopic metallic isolated networks. We calculate the average and typical
magnetizations in the diffusive regime for non-interacting electrons
or in the first order Hartree-Fock approximation. These
quantities are related to
the return probability for a diffusive particle on the corresponding network.
By solution of the diffusion equation on various types of networks,
including a ring with arms or an infinite square network,
we deduce the corresponding magnetizations.
	In the case of an infinite network, the Hartree-Fock average
 magnetization stays finite in the thermodynamic limit.

} \end{abstract}
\end{center}
]
\narrowtext
The problem of persistent currents in mesoscopic
 rings\cite{Bloch65,Kulik70,Buttiker83} has been
stimulated by a few key experiments in the recent
years.
Two types of measurements have been deviced, single ring
experiments\cite{Chandrasekhar91,Mailly93,Webb95} and
many
rings experiments\cite{Levy90,Reulet95,Webb95}. In the second case, the measured
quantity is
an average magnetization $\langle M \rangle$ while the first type of
experiment can only give the
magnetization corresponding to a given disorder configuration. In the last
case, the width $M_{typ}$ of the magnetization distribution is 
also of interest:
$M_{typ}^2 = \langle M^2 \rangle - \langle M \rangle^2$.

Essentially two types of methods have been used: i) analytical methods
where the diffusive electronic motion is treated in a
perturbative way, leading to the famous Cooperon diagrams; non-interacting
electron 
theory\cite{Cheung89,Schmid91,VonOppen91,Altshuler91,Akkermans91a,Oh91,Argaman93a}
or Hartree-Fock approximation
\cite{Schmid91,Ambegaokar90,Eckern91,Argaman93b,Montambaux96}
have been considered;
ii) strictly 1D models or numerical methods in which either there
is no  diffusive
motion or the system size is too small to give
quantitative results\cite{interactions}.
Up to now, the only 
 results for the diffusive regime are given by the perturbative 
method.  The status of the comparison
 with experiments is not completely clear yet but its 
 seems to be a reasonable agreement 
with preliminary recent  data\cite{Webb95}, the typical
 current being properly
well described by the non-interacting 
theory\cite{Cheung89,Argaman93a,Montambaux95} and the average current being
described by the Hartree-Fock
 approximation\cite{Ambegaokar90,Eckern91,Argaman93b,Montambaux96,Montambaux95}.

We propose that a new  way to get insight into the problem is
to study {\it other
geometries} than simple rings. In this letter, we calculate analytically the
typical and average magnetizations of various types of networks, following
the method (i).
To do so, we use a semi-classical picture to  relate the quantities of
interest to the return probability of a classically diffusive particle.
Then, this return probability is calculated on different type of networks,
giving access to the magnetization.
As examples, we treat the case of an isolated ring connected to one or two
arms, and the case of an infinite square lattice.
Several experiments are
proposed (otherwise specified, $\hbar=1$ throughout the paper).

In the absence of $e-e$ interactions, a finite
 contribution to the average magnetization comes from the
fact that the number $N$ of particles is fixed in each subsystem of the
ensemble\cite{Bouchiat89}. It turns out that this contribution
is by far smaller than the experimental results. However we will discuss
it mainly for pedagogial purpose and comparison with other contributions.
With this constraint on $N$, the "canonical"magnetization
is given by\cite{Imry89}: \begin{equation}
\langle M_N(H)\rangle=-{\Delta \over 2}{\partial \over \partial H}
\langle\delta N^2(\mu)\rangle\,,
\label{MN}
\end{equation}
where $\Delta$ is the mean level spacing and $\langle \delta N^2(\mu)
\rangle$ is the sample to sample fluctuation
of the number of single-particle states below the Fermi energy $\mu$. It is
an integral of the two-point correlation function of the
density of states (DOS) $K(\varepsilon_1
- \varepsilon_2)= \langle \rho(\varepsilon_1)\rho(\varepsilon_2) \rangle
-\rho_0^2$ . $\rho_0$ is the average DOS. $K(\varepsilon)$ has been
calculated by Altshuler
and Shklovskii\cite{Altshuler86} and later in the presence of a magnetic
flux\cite{Schmid91,VonOppen91,Altshuler91}.
A very useful semiclassical
picture has been presented by Argaman {\it et al.}, which
relates the Fourier transform $\tilde K(t)$ of
$K(\varepsilon)$ to the classical return probability $p(\r,\r,t)$ for a
diffusive particle\cite{Argaman93a}.

\begin{equation} \label{AIS}
\tilde K(t) = {t P(t) \over 4 \pi^2} \end{equation}
where $P(t) = \int p(\r,\r,t) d\r$. This return probability has two
components, the purely classical one and the interference term which results
from interferences between pairs of time-reversed trajectories. In the
diagrammatic picture, they are related to the diffuson and Cooperon diagrams.
The interference term is field dependent and is solution of  the
 diffusion equation:

\begin{equation}  \label{diff}
[{\partial \over \partial t} -  D(\nabla + {2 i e {\bf A} \over \hbar
c})^2]p(\r,\r',t)= \delta(t) \delta(\r-\r')
 \end{equation}
From eqs. (\ref{MN},\ref{AIS}), the average canonical 
magnetization can be related to
the field dependent part of the return probability:

\begin{equation} \label{MN2}
\langle M_N(H)\rangle=-{\Delta \over 4 \pi^2 }{\partial \over \partial
H} \int_0^\infty {P(t,H)\over t} dt\,,
\end{equation}
Note that the field dependent part of this integral converges at
small times. At large times, the return probability is
exponentially cut-off as $e^{-\gamma t}$ where $\gamma =\hbar D
/L_\varphi^2$ is the inelastic scattering rate.

Due to the $e-e$ interactions, a
 larger  contribution to the average magnetization exists, 
which has been calculated by
Ambegaokar and Eckern\cite{Ambegaokar90}, in the Hartree-Fock
approximation.
It can be written
as\cite{Ambegaokar90,Schmid91,Argaman93b,Montambaux96}:
\begin{equation} \label{Mee}
\langle M_{ee}(H)\rangle =-{U \over 4} {\partial \over \partial H}\int
\langle n(\r)^2 \rangle
d\r
\end{equation}
Where $U$ is an effective screened interaction and 
$n(\r)$ is the local density. 
The integrand is related to the fluctuations of the local DOS which in
 turn
can be related to the return probability\cite{Montambaux96}. One
gets:
\begin{equation}
\label{Mee2}
\langle M_{ee}(H)\rangle =- {U\rho _0 \over \pi }{\frac
\partial {\partial H }}\int_0^\infty {\frac{P(t,\phi )}{t^2}}dt
\end{equation}

In a similar way, the
typical magnetization  can also be straightforwardly written in terms of
$K(\varepsilon)$\cite{Argaman93a,Montambaux95}. By Fourier
transform, one has:

\begin{equation}  \label{Mtyp}
M_{typ}^2(H)={1 \over 8 \pi^2}
\int_0^\infty {P"(t,H)|^H_0 \over t^3} dt \,,
\end{equation}
where $P"(t,H)|^H_0=   \partial^2 P/\partial H^2|_0 -   \partial^2
P/\partial H^2|_H$.

To be complete, we remind that the weak-localization correction to the
conductance of a connected mesocopic sample  can be also be related to
the return probability\cite{Khmelnitskii84,Doucot86}:
\begin{equation} \label{Wloc}
\Delta \sigma(r) =(-2/\pi \rho_0)  \sigma_0 C_\gamma(\r,\r)
\end{equation}
$\sigma_0$ is the Drude conductivity. The Cooperon $C_\gamma(\r,\r,H)$
 is the time integrated field-dependent return probability:
\begin{equation} \label{Cgamma}
 C_\gamma(\r,\r,H) = \int_0^\infty p(r,r,t,H) dt
\end{equation}
It appears that all the quantities of interest are obtained as time integrals
of the return probability with various power-law weighting functions.
Noting that $P(t)$ has the form $P_0(t)e^{-\gamma t}$ and that
\begin{equation}  \label{PC}
\int {P_0(t) \over t } e^{-\gamma t} dt =  \int_\gamma^\infty d\gamma \int
  C_\gamma(\r,\r,H) d\r
\end{equation}
the different magnetizations can be given in terms of the successive
integrals of $C_\gamma(\r,\r,H)$: \begin{eqnarray}  \label{MN3}
\langle M_N(H)\rangle &=& -{\Delta \over 4 \pi^2}{\partial \over \partial H}
\int C_\gamma^{(1)}(\r,\r,H) d{\r}   \\
\label{Mee3}  \langle M_{ee}(H)\rangle &=& -{U\rho_0 \over \pi}{\partial
\over \partial H} \int C_\gamma^{(2)}(\r,\r,H) d{\r}   \\
\label{Mtyp3} M_{typ}^2(H) &=& {1 \over 8 \pi^2}{\partial^2 \over \partial
H^2} \int C_\gamma^{(3)}(\r,\r,H)  \big|^H_0 d\r  
\end{eqnarray}
where    $C^{(n)}_\gamma = \int_\gamma^\infty d\gamma_n ...
\int_{\gamma_2}^\infty d\gamma_1 \int_{\gamma_1}^\infty d\gamma'
C_{\gamma'} $. These are the key equations
of this paper since the different magnetizations can be calculated from the knowledge
 of 
the return probability $C_\gamma(\r,\r,H)$ on the different
lattices considered and can be deduced from each other or related to
weak-localization correction by  $H-$ or $\gamma-$ derivatives or
integrations.

For the case of weak-localization correction, an extensive study of this
quantity on
various lattices has been carried out by Dou\c cot and Rammal\cite{Doucot86}.
Considering networks made of quasi-1D wires so that the diffusion can be
considered as one-dimensional, the
Cooperon $C_\gamma(r,r')$ obeys the diffusion equation
\begin{equation}   \label{diff2}
[\gamma - \hbar D (\nabla + {2 i e A \over \hbar c})^2]
C_\gamma(r,r')=\delta(r-r')
\end{equation}
with the continuity equations
\begin{equation}  \label{cal}
\sum_\beta ( -i {\partial \over \partial r} - {2 e A \over \hbar c})_\alpha
C_\gamma(\alpha,r') = {i \over D S }
\delta_{r',\alpha}
\end{equation}
$r,r'$ are  linear coordinates on the network and $\alpha,\beta$ are
 nodes. The term $\gamma$ in eq. \ref{diff2} describes the inelastic
scattering.
Integration of the differential equation (\ref{diff2}) with the boundary
conditions (\ref{cal}) leads to the so-called network equations
which relate $C_\gamma(\alpha,r')$  to the
neighboring nodes $\beta$\cite{Doucot86}.
$$
\sum_\beta\coth({l_{\alpha\beta} \over L_\varphi})C(\alpha,r')-
\sum_\beta{C(\beta,r')e^{-i\gamma_{\alpha\beta}}\over\sinh(l_{\alpha\beta}/L_\varphi)}
={L_\varphi \over D S} \delta_{\alpha,r'}   
$$
$l_{\alpha\beta}$ is the length of the link $(\alpha\beta)$ and 
$\gamma_{\alpha\beta}=(4 \pi /\phi_0)\int_\alpha^\beta {\bf A} d {\bf l}$
 is the circulation of the vector potential along this link.
 Then $C_\gamma(r',r')$ is calculated in terms of the $C_\gamma(\alpha,r')$.
Finally, spatial integration give access to the magnetizations.

As an example, we have first considered the case of a ring of perimeter $L$
connected to
an arm of length $b$.
Such a geometry has been considered in the stricly
 1D case without disorder\cite{Buttiker85,Akkermans91b,Mello93}.
 It is expected that since the electrons will spend some
time in the arm where there are not sensitive to the flux, the persistent
current will be decreased.
From eqs. \ref{diff2},\ref{cal}, the function
$C_\gamma(r,r)$ can be straightforwardly calculated
on the arm and on the
ring. Spatial integration gives:
$$
\langle M_N \rangle = {\Delta S\over \pi \phi_0} {\sin 4 \pi
\varphi
\over {1 \over 2} \tanh {b\over L_\varphi} \sinh{L\over L_\varphi}+ \cosh
{L\over L_\varphi} - \cos 4 \pi \varphi }
$$
where $\varphi = \phi/\phi_0$. $\phi$  is the flux through the ring,
$\phi_0 =h/e$ is the flux quantum and $S$ is the area of the ring.
 Writing this magnetization as $\langle
M_N \rangle = \sum_m \langle M_N \rangle_m \sin 4 \pi m \varphi$, the
harmonics content is given by
\begin{equation}   \label{MNm1}
\langle M_N \rangle_m = {2 \Delta S \over \pi \phi_0}
e^{- m \arg\cosh[\cosh{L \over L_\varphi} + {1 \over 2} \tanh{b
\over L_\varphi} \sinh{L \over L_\varphi}]}\end{equation}
Well-known results are recovered when $b=0$.
In the case where $L \approx L_\varphi$, the harmonics content  can be
simply written as:

\begin{equation}   \label{MNm2}
\langle M_N \rangle_m = {2 \Delta S\over \pi \phi_0}
({2 \over 2 +\tanh b /L_\varphi})^m e^{- m L / L_\varphi}
\end{equation}
so that in the limit $b \rightarrow \infty$, the harmonics are reduced by
a ratio $(2/3)^m$, compared to the ring. 
However the
arm has another much more dramatic effect which is to decrease the
interlevel spacing:
 $\Delta(b) = \Delta(0) L /(L + b)$. This reduction does not exist for
$\langle M_{ee} \rangle$ and $M_{typ}$ which can
also
be simply calculated by successive integrations on $\gamma$. Here we give
the result for an arm of length $b \gg L_\varphi$:
 \begin{equation}
 \langle M_{ee}(\infty) \rangle_m = ({2 \over 3})^m
 \langle M_{ee}(0) \rangle_m
\end{equation}
\begin{equation}
  M_{typ}(\infty)_m = ({2 \over 3 })^{m/2}  M_{typ}(0)_m
\end{equation}
where $\langle M_{ee}(0) \rangle_m$ are the magnetizations of the 
isolated loop.
 Interestingly, it is seen that the magnetizations  $\langle M_{ee} \rangle$
and  $ M_{typ}$ do not decrease to $0$ when $b \rightarrow \infty$ but
{\it saturate} to finite values
with respective reductions of the first harmonics in the ratios  $2/3$ et
${\sqrt{2/3}}$. In the limit $b \gg L_\varphi$, the magnetization
 should be unchanged if a reservoir is attached to the arm\cite{Buttiker85}.
The case of a ring connected to two arms of length $b$ shown in
figure 1 can be treated
in a very similar way.  In this case we find that the lowest harmonics
of the $e-e$ average current is reduced in a ratio  4/9 and the typical
current is reduced by a factor 2/3.
We propose that single ring experiments with appropriately designed
arms could be able to measure these reductions.
\begin{figure}
\centerline{
\epsfxsize 6cm
\epsffile{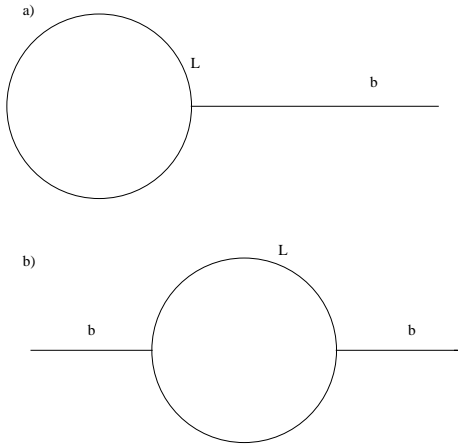}}
\caption{Two geometries considered in this letter
}
\label{fig1}
\end{figure}
We now turn to the case of an infinite square lattice
whose
magnetization will be compared with the one of an array of isolated rings.
 The eigenvalues of the diffusion equation can
be calculated for a rational flux per plaquette $\varphi=\phi/\phi_0=p/2q$.
Denoting by $a$ the lattice parameter, $\eta = a/L_\varphi$ and  $\phi=H
a^2$ the flux per plaquette, we
find that the canonical magnetization per plaquette is
\begin{equation}  \label{MNnetwork}
\langle M_N \rangle = {\Delta \over 4 \pi^2 q} {\partial
\over \partial H} \sum_{i=1}^q \{ \ln(4 \cosh \eta
-\varepsilon_i(\theta,\mu))\} \end{equation}
where
$\{ \ldots \} = \int_0^{2\pi} {d\theta  \over 2 \pi}
\int_0^{2\pi} {d\mu \over 2 \pi} \ldots $. $\varepsilon_i(\theta,\mu)$ are
the solutions of the determinental equation
\begin{equation}  \label{Matrix}
det M = det\left|\begin{array}{cccccc}
M_1           & 1    &  0    &\ldots  &0     &e^{i\mu}      \\
1             &M_2   &      1&\ldots  & 0    &0                  \\
\vdots        &\vdots&       &\ddots  &      &\vdots                  \\
e^{-i\mu}  &0     &0      &\ldots  &1     &M_q        \\
\end{array}    \right| =0
\end{equation}
where $M_n = 2 \cos(4 n\pi \varphi+\theta/q) - \varepsilon$. $M$ is a matrix
associated to the Harper equation known to be also relevant for other related
problems like tight-binding electrons in a magnetic field\cite{Hofstadter76}
or superconducting networks in a field\cite{Rammal83}.

The magnetization per plaquette can be compared to the magnetization of a 
square 
ring of perimeter $L=4 a$:
\begin{equation}  \label{MNring}
\langle M_N \rangle = {\Delta \over 4 \pi^2 } {\partial
\over \partial H} \ln( \cosh 4 \eta -\cos 4 \pi \varphi)
\end{equation}
Since $\Delta \rightarrow 0$ for the infinite network,
this canonical magnetization density vanishes for an infinite
network as it was already noticed for a chain of connected
rings\cite{Ketteman95}. On the other  hand, the $e-e$ contribution stays
finite in the thermodynamic limit. It is given by:
\begin{equation} \label{Meenetwork}
\langle M_{ee} \rangle = U \rho_0{e D \over \pi^2 q} {\partial
\over \partial \varphi } \sum_{i=1}^q \int_\eta^\infty \{ \ln(4 \cosh \eta
-\varepsilon_i(\theta,\mu))\} \eta d \eta 
\end{equation} and  can be compared with
 the ring magnetization which can be cast in the form:
\begin{equation}  \label{Meering}
\langle M_{ee} \rangle =  U \rho_0{4 e D \over \pi }
\int_\eta^\infty {\sin 4 \pi \varphi \over  \cosh 4 \eta -cos 4 \pi \varphi}
\eta d\eta  \end{equation}
(This integral can be calculated explicitely in terms of the
 Lobatchevsky function and it has the Fourier decomposition found by 
Ambegaokar and Eckern\cite{Ambegaokar90}).
Contrary to the canonical magnetization, the $e-e$ magnetization
is an extensive quantity.
This magnetization density is plotted on figure 2 for the ring and the
infinite lattice. It is first 
seen that the network magnetization is continuous.
Although the field dependence of the eigenvalues of the Harper equation
has a very complicated discontinuous behavior ( the so-called Hofstadter
spectrum), the sum on the eigenvalues has a smooth behavior\cite{Tan92}.
$\langle M_{ee} \rangle$    can be calculated easily for large $q$. In this
 case the dispersion $\varepsilon_i(\theta,\mu)$ is very small and the density
of states can be approximated by a
 sum of $\delta$ functions\cite{Montambaux89}.
 A very good approximation of the sum (\ref{Meenetwork}) can be
 obtained by replacing the integrals on $\theta$ and $\mu$
by the value of each term in the sum taken at $\varepsilon_i(\pi,\pi)$.
 \begin{figure}
\centerline{
\epsfxsize 8cm
\epsffile{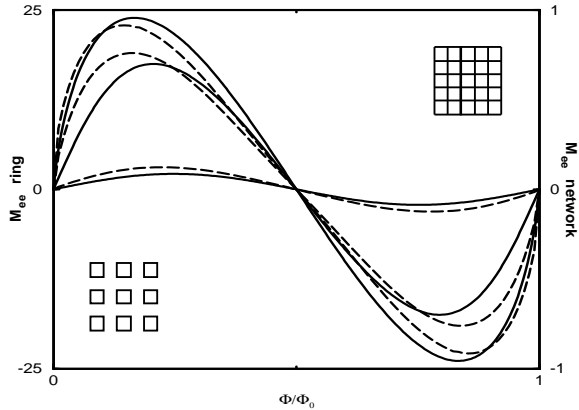}}
\caption{Average magnetization $\langle M_{ee} \rangle$ of a single ring (full lines)  and
magnetization
density of the infinite network (dashed lines), for $L_\varphi=\infty$, $4 a$
and $a$ } \label{fig2}
\end{figure}

It is seen on figure 2 that the network magnetization density
is about 25 five times smaller than the 
ring magnetization. 
Considering that on the array of square rings already considered
experimentally\cite{Levy90}, the distance between rings is
equal to the size of the squares,
the number of squares is four times larger when they are connected.
One then expect only a factor of order $6$ between the 
magnetization of the array of disconnected rings and the lattice.
The width of the magnetization distribution scales as $1/\sqrt{S}$, $S$
being the area of the network.

Our results have been obtained in the 
Hartree-Fock approximation and should be corrected by higher order
contributions\cite{Eckern91,Altshuler85}. But we do not see any reason why the
ratio between these two magnetizations could be drastically changed.
Thus we propose that the measurement of the magnetization of a mesoscopic
network should give  similar results to 
the one of an array of disconnected rings.

G.M. thanks the Institute of Theoretical Physics, UCSB,
 for hospitality during the completion of this work.
 This research was supported in part by the
NSF Grant No. PHY94-07194.

\end{document}